\documentclass[aps,prl,reprint,showpacs]{revtex4-1}

\usepackage[dvipdfmx]{graphicx}
\usepackage{mediabb}
\usepackage{amsmath,amssymb}
\usepackage{color}
\usepackage{bm}
\usepackage{ulem}

\def\v#1{\mathbf{#1}}			

\def\vq{\v{q}} 					
\def\vk{\v{k}} 					

\def\r{\v{r}} 					
\def\q{\v{q}} 					
\def\k{\v{k}} 					

\def\mat#1#2#3#4{\left(
\begin{array}{cc} #1 & #2 \\#3& #4\end{array} \right)}

\def\vvv#1#2{\left(
\begin{array}{c} #1  \\#2\end{array} \right)}

\def\del{\partial}

\def\la{\langle}
\def\ra{\rangle}

\def\tsf{\tau_{\rm sf}}

\def\IM{{\rm Im}} 
\def\ikqom{\int_{\vq\vk\omega}} 

\def\TR{{\rm Tr}}

\def\IsSSE{I_S^{\rm SSE}}
\def\IsSP{I_S^{\rm SP}}
\def\Istot{I_S^{\rm tot}}
\def\SSSE{\mathcal{S}^{\rm SSE}}
\def\SSP{\mathcal{S}^{\rm SP}}
\def\Seq{\mathcal{S}^{\rm eq}}
\def\Stot{\mathcal{S}^{\rm tot}}
\def\FSSE{\mathcal{F}^{\rm SSE}_S}
\def\FSP{\mathcal{F}^{\rm SP}_S}

\begin{document}


\title{Spin current noise of the spin Seebeck effect and spin pumping}


\author{M. Matsuo$^{1,2}$, Y. Ohnuma$^{2}$, T. Kato$^{3}$ and S. Maekawa$^{2}$}
\affiliation{%
${^1}$Advanced Institute for Materials Research, Tohoku University, Sendai, 980-8577, Japan.\\
${^2}$Advanced Science Research Center, Japan Atomic Energy Agency, Tokai 319-1195, Japan.\\
${^3}$Institute for Solid State Physics, University of Tokyo, Kashiwa, 277-8581, Japan.
}%

\date{\today}

\begin{abstract}
We theoretically investigate the fluctuation of a pure spin current induced by the spin Seebeck effect and spin pumping 
in a normal metal (NM)/ferromagnet (FM) bilayer system.
Starting with  a simple FI--NM interface model with both spin-conserving and spin-non-conserving processes, 
we derive general expressions of the spin current and the spin-current noise at the interface within second-order perturbation of
the FI--NM coupling strength, and estimate them for an  yttrium iron garnet (YIG) --platinum interface.
We show that the spin-current noise can be used to determine the effective spin 
carried by a magnon
modified by the spin-non-conserving process at the interface. 
In addition, we show that it provides information on the effective spin of a magnon, heating at the interface under spin pumping, and spin Hall angle of the NM.
\end{abstract}

\pacs{72.20.Pa, 72.25.-b, 85.75.-d}

\maketitle 

\paragraph{Introduction.---}
In mesoscopic physics, it is well known that measurement of electrical current noise 
through a device provides useful information about electron transport~\cite{Blanter00,Martin05}.
Equilibrium noise or Johnson--Nyquist noise~\cite{Johnson28,Nyquist28} is related
to effective electron temperatures in a device according to the fluctuation--dissipation theorem~\cite{Onsager31,Casimir45,Green54,Kubo57}.
Nonequilibrium current noise under a high voltage bias, for example, shot noise~\cite{Schottky18}
can be used for determining the effective charge of a quasi-particle~\cite{Picciotto97,Saminadayar97,Jehl00,Kozhevnikov00,Zarchin08, Ferrier16},
direct demonstration of Fermi statistics of electrons~\cite{Reznikov95,Kumar96,Buttiker90,Landauer91,Buttiker92,Martin92},
and evaluating nonequilibrium spin accumulation~\cite{Arakawa2015,Iwakiri17}.

As expected from fruitful physics of the current noise, 
fluctuation of the pure spin current, that is, {\it spin-current noise} has a potential to provide important information on spin transport in a spintronics device.
Spin-current noise has been measured by converting it into the voltage noise 
induced by the inverse spin Hall effect, and has been used to obtain information about spin transport within the fluctuation dissipation relation regime~\cite{Kamra14}.
Recently, spin-current noise of spin pumping as well as equilibrium noise has been studied theoretically
~\cite{Kamra16a,Kamra16b}.
Spin-current noise of spin Seebeck effect has been discussed for
one-dimensional spin chains~\cite{Aftergood17}. 
These works employ simple microscopic models with spin-conserving exchange interactions, and put a special emphasis on exotic properties characteristic of specific systems.
Spin-current noise has, however, not been utilized so far to access microscopic information, which addresses the important problems in the field of spintronics such as separation of a spin current according to the driving mechanism.

\begin{figure}[!tb]
\begin{center}
\includegraphics[scale=0.4]{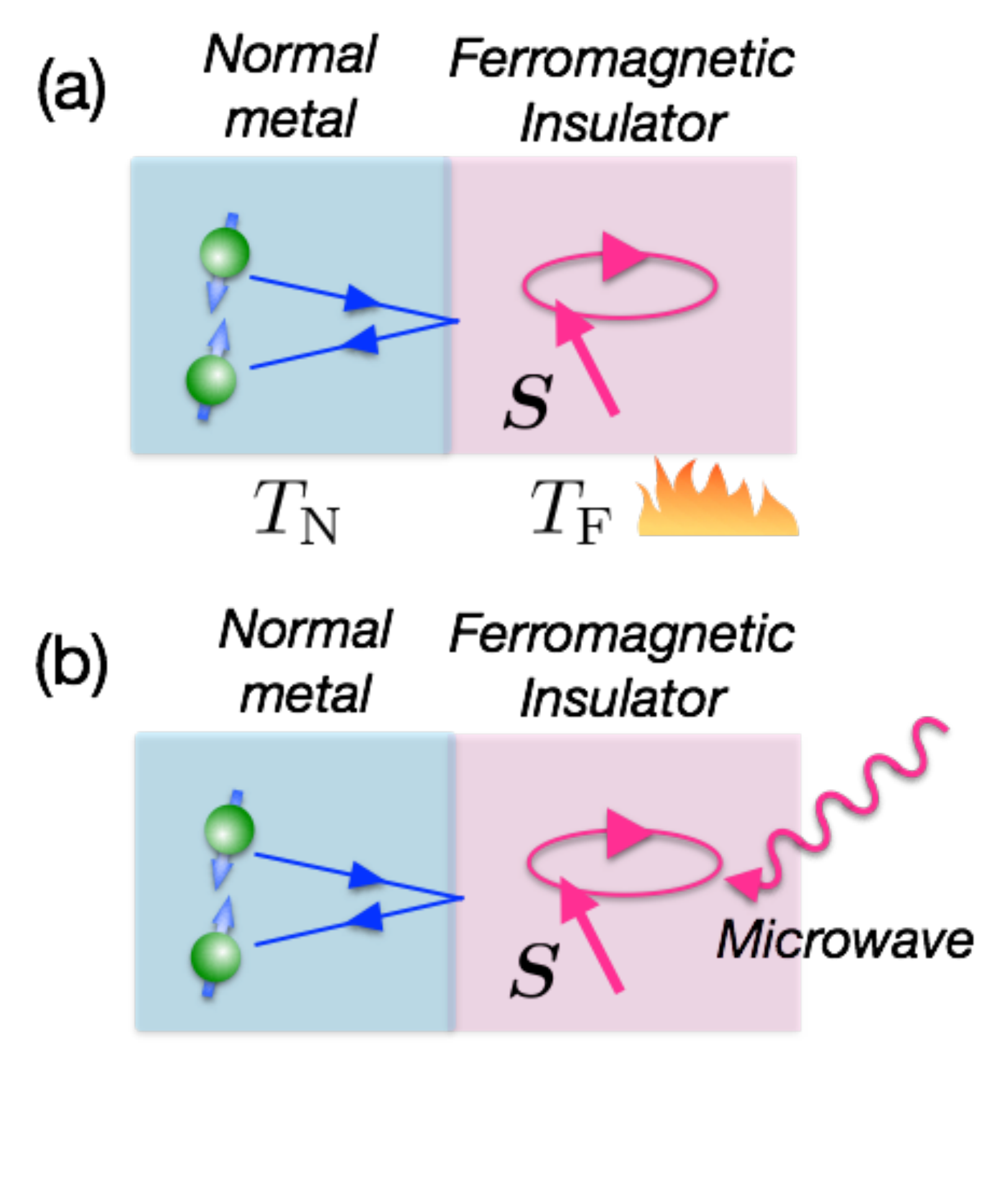}
\caption{(Color online) Two mechanisms for spin-current generation at the interface between a normal metal and a ferromagnetic insulator. (a) Spin Seebeck effect driven by temperature biases ($T_N \ne T_F$) and (b) spin pumping generated by ferromagnetic resonance (i.e., irradiation with microwaves). }
\label{fig_setup}
\end{center}
\end{figure}

In the field of spintronics, the spin current through the interface between a normal metal (NM) and a ferromagnet insulator (FI) 
is a central issue in many experiments~\cite{MaekawaEd2012}.
For example, spin current flows through the interface according to the spin Seebeck effect 
in the presence of a temperature difference between NM and FI~\cite{Uchida08,Jaworski10,Uchida10,Xiao10,Adachi11,Ohnuma17} (see Fig.~\ref{fig_setup}~(a)).
Moreover, spin current is produced by ferromagnetic resonance (FMR), which is achieved by irradiating FI with microwaves~\cite{Tserkovnyak02,Konig03,Saitoh06,Kajiwara10,Ohnuma14} (see Fig.~\ref{fig_setup}~(b)).
The generation of spin current in these two setups is important in many spintronics applications using metallic materials.
Therefore, whether new information is obtained by measuring the spin-current noise in an NM--FI bilayer system is a fundamental question.

In this study, we theoretically investigate the spin-current noise at the FI--NM interface, and
show that spin-current noise provides useful information about spin transport.
Starting with a microscopic model of the FI--NM interface, we derive general expressions for spin current and spin-current noise within the framework of Keldysh Green's function~\cite{Haug-text}, and estimate them for an yttrium iron garnet (YIG)--platinum interface.
At sufficiently low temperatures, the spin-current noise becomes independent of the temperature (spin shot noise), and includes information about an effective magnon spin determined by the ratio of the spin-conserving process to the spin-non-conserving process.
In addition, we show that measurement of the spin current noise provides useful information about the heating effect under spin pumping 
and the spin Hall angle of the NM.

\begin{figure}[!tb]
\begin{center}
\includegraphics[scale=0.45]{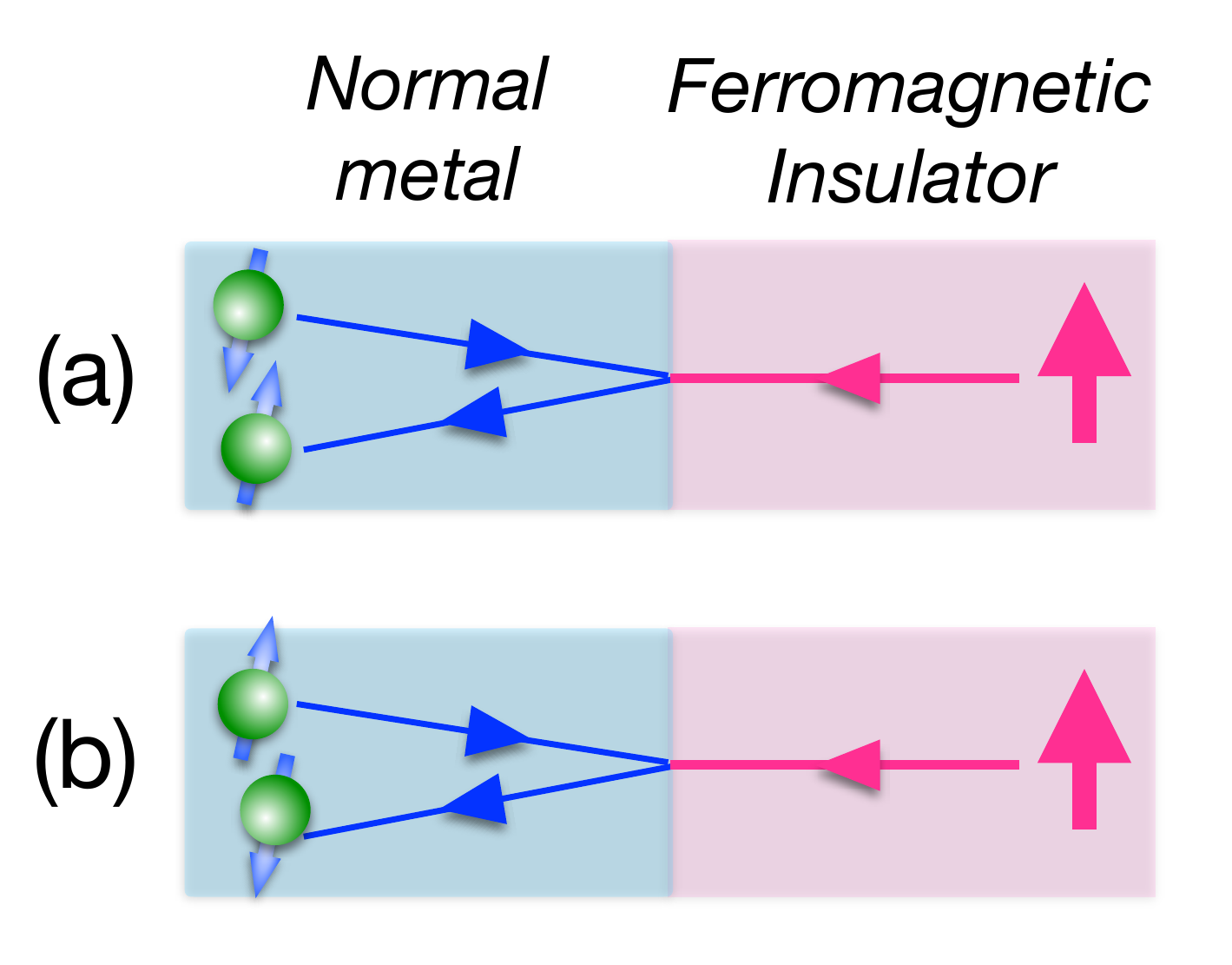}
\caption{(Color online) Two types of spin conversion at interface. (a) The spin--conserving process is described by $H_1$, and (b) the spin--non-conserving process is described by $H_2$. }
\label{fig_spinconversion}
\end{center}
\end{figure}

\paragraph{Model.---}
Consider spin transport in a bilayer system, where a NM and a ferromagnet (FM)  interact through $s$-$d$ exchange at the interface (see Fig.~\ref{fig_setup}). 
The NM is described by non-interacting conduction electrons, whereas the FI is by the Heisenberg model with Zeeman energy $H_Z=\sum_i S_i^z \gamma h_0$, where $S_i$ represent the localized spin in the FM,
$\gamma$ is the gyromagnetic ratio, and $h_0$ denotes an external magnetic field.
The interface is modeled using the Hamiltonian,  (see Supplemental Material for details)
$H = H_1 + H_2 $:
\begin{eqnarray}
H_1 &=& \sum_i (J^z_1\sigma^z_i S^z + J_1\sigma^+_i S^-_i  + J_1^*\sigma^-_i S^+_i), \\
H_2 &=& \sum_i (J_2\sigma^+_i S^+_i  + J_2^*\sigma^-_i S^-_i),
\end{eqnarray}
where $\sigma_i$ represents the conduction electron spin in the NM.
In addition to the interfacial exchange interaction $H_1$, which conserves the spin angular momentum, we consider the spin non-conserving interaction described by $H_2$ (see ).
Figure~\ref{fig_spinconversion}~(a) and (b) indicate the processes described by $H_1$ and $H_2$, respectively. 
The present interface model can be derived generally by assuming the presence of anisotropic exchange interaction or magnetic dipole-dipole interaction. 

\paragraph{Spin current.---}
The spin current generated at the interface can be calculated as the rate of change of conduction electron spin in the NM, $\la \hat{I}_S \ra := \hbar \sum_i \la \del_t \sigma^z_i \ra $, where $\hat{I}_S := \hbar \sum_i  \del_t \sigma^z_i $ is a spin current operator and $\la \cdots \ra := \TR [ \hat{\rho} \cdots ]$ denotes the statistical average with the density matrix $\hat{\rho}$.
The spin current operator is expressed by a sum of the components: $\hat{I}_S = \sum_{a=1}^2 \hat{I}_S^a$, where $\hat{I}_S^a := (i\hbar)^{-1}\sum_i [\sigma_i^z, H_a]$. $(S_i^1, S_i^2) := ( S_i^-, S_i^+)$. 
The second-order perturbation with respect to the interfacial interactions and the spin wave approximation at the lowest order of $1/S$ expansion yield the following expression (see Supplemental Material for details):
\begin{eqnarray}
\la \hat{I}^a_{S}\ra  &=&  2A_a \!\! \ikqom \!\! \IM \chi^{R}_{\q\omega} \IM G^{R,a}_{\k\omega} \delta f^{{\rm neq},a}_{\r t\k\q\omega}, \label{Is-a}
\end{eqnarray}
where $A_a = 4N_{\rm int}^a J_a{}^2/\hbar $, $\chi^{R}$ is the spin susceptibility, $G^{R}$ is the retarded Green's function for localized spin, 
and a random average is taken over the impurity positions at the interface. 
The nonequilibrium distribution difference between PM and FI is defined as $\delta f^{{\rm neq},a}_{\r t\k\q\omega} = f^{N}_{\r t\q\omega}-f^{F,a}_{\r t\k\omega}$~\cite{footnote1,Du17}, and the relations $\IM \chi^{R}_{\q,-\omega}= -\IM \chi^{R}_{\q\omega}$ are used. 
Formula (\ref{Is-a}) is regarded as a counterpart of the well-known formula for tunnel junctions, which are described by combinations of densities of states and the difference in Fermi distribution functions between two normal metals~\cite{Ingold92,Bruus04}.

\paragraph{Spin-current noise.---}
We introduce the spin-current noise as follows:
\begin{eqnarray}
\mathcal{S}_{t_1 t_2} = \frac12 \la  \{ \hat{I}_{S}(t_1), \hat{I}_{S}(t_2) \} \ra,
\end{eqnarray}
where $\{ \hat{I}_{S}(t_1), \hat{I}_{S}(t_2) \} = \hat{I}_{S}(t_1) \hat{I}_{S}(t_2)+\hat{I}_{S}(t_2) \hat{I}_{S}(t_1)$. In addition, we introduce the dc-limit of noise power $\mathcal{S}:=\mathcal{S}(\omega=0)$, where $\mathcal{S}(\omega) := T^{-1}\int_0^T dt_1 \int_0^T dt_2 e^{i\omega (t_1-t_2)} \mathcal{S}_{t_1 t_2}$. 
The noise power consists of equilibrium and nonequilibrium parts as $\mathcal{S}^{\rm tot} = \mathcal{S}^{\rm eq} +\mathcal{S}^{\rm neq}$, and the second-order perturbation with respect to the interfacial interactions yields the following expression:
\begin{eqnarray}
&&\!\!\!\! \mathcal{S}^{\rm eq}=2\hbar ( A_1\! +\! A_2) \!\!\int_{\k\q\omega} \!\!\!\!\!\!\!\! \IM \chi^{R}_{\q\omega} \IM G^{R}_{\k\omega}f^N_\omega(1+f^N_\omega), \label{eq:Seq}\\
&&\!\!\!\! \mathcal{S}^{\rm neq}\!\!=\hbar ( A_1\! +\! A_2) \!\!\int_{\k\q\omega} \!\!\!\!\!\!\!\! \IM \chi^{R}_{\q\omega} \IM G^{R}_{\k\omega}  (1+2f^N_\omega) \delta f^{{\rm neq}}_{\r t\k\q\omega}. \label{eq:Sneq}
\end{eqnarray}
Here, we have abbreviated $f^N_{\r t\q\omega}$ as $ f^N_\omega$, and
$\delta f^{{\rm neq}}_{\r t\k\q\omega} = f^N_{\r t\q\omega} - f^{F}_{\r t\k\omega}$
with $f^{F}_{\r t\k\omega} := G^{<,1}_{\k\r t\omega}/2i \IM G^{R,1}_{\k\omega}$.
We note that $\mathcal{S}^{\rm eq}$ is independent of the nonequilibrium distribution difference, and is determined purely by the distribution functions in thermal equilibrium.

\paragraph{Spin Seebeck effect.---}
Regarding the spin Seebeck effect, the distribution difference
$\delta f^{{\rm neq}}_{\r t\k\q\omega}$
is caused by the temperature bias at the interface as $f^N_{\r t\q\omega} =f^0_\omega(T_N)$ and $f^{F,1}_{\r t\k\omega} =f^0_\omega(T_F)$, $f^{F,2}_{\r t\k,-\omega} =f^0_{-\omega}(T_F)$, and $\IM G^{R,2}_{\k\omega} = - \IM G^{R}_{\k,-\omega}$ with $f^0_\omega(T)= (e^{\hbar \omega/k_B T}-1)^{-1}$. 
By substituting these distribution functions into Eqs.~(\ref{Is-a}) and (\ref{eq:Sneq}),  
we obtain the spin--Seebeck current $I_S^{\rm SSE}$ and the spin--Seebeck noise $\mathcal{S}^{\rm SSE}$:
\begin{eqnarray}
& &  \!\!\!\!  I_S^{\rm SSE}
=
  (A_1\! -\! A_2) \!\! \ikqom \!\!\!\!\!\! \IM \chi^{R}_{\q\omega} \IM G^{R}_{\k\omega} \delta f^{\rm SSE}_{\k \omega} , \label{IsSSE} \\
& & \!\!\!\! \mathcal{S}^{\rm SSE}\!\!=\hbar ( A_1\! +\! A_2) \!\!\int_{\k\q\omega} \!\!\!\!\!\!\!\! \IM \chi^{R}_{\q\omega} \IM G^{R}_{\k\omega} (1+2f^N_\omega) \delta f^{\rm SSE}_{\k \omega} , \label{eq:SneqSSE}
\end{eqnarray}
where $\delta f^{\rm SSE}_{\k \omega} = \frac{\del f^0_\omega}{\del T} \Delta T$ and $\Delta T = T_N -T_F$. 

\paragraph{Spin pumping.---}
Next, we consider spin-current noise in the case of spin pumping. 
Here heating at the interface by using microwave irradiation is neglected for simplicity. 
The nonequilibrium source used to generate the spin current is the ferromagnetic resonance in the FM, and it is described by the Hamiltonian $H_{\rm ac} = \frac{\hbar \gamma h_{\rm ac}}{2}(S^+ e^{i\Omega t} + S^- e^{-i\Omega t})$, where $h_{\rm ac}$ and $\Omega$ are the amplitude and frequency of the microwaves, respectively.
Substituting 
$f^{N}_{\q\omega}-f^{F,1}_{\k\omega} = \delta f^{\rm SP}_{\k\omega}(\Omega)$ and 
$f^{N}_{\q\omega}-f^{F,2}_{\k\omega} = \delta f^{\rm SP}_{\k\omega}(-\Omega)$, 
where $\delta f^{\rm SP}_{\k\omega} (\Omega)= 2S_0(\gamma h_{\rm ac}/2)^2N_F\delta_{\k 0} \pi \delta (\omega -\Omega)/\alpha \omega$, 
into Eqs.~(\ref{Is-a}) and (\ref{eq:Sneq}), we obtain the following expressions:  
\begin{eqnarray}
&& I_S^{\rm SP} = (A_1 \! - \! A_2) g(\Omega) \int_{\q} \IM \chi^R_{\q \Omega}, \label{eq:IsSP} \\
&& \mathcal{S}^{\rm SP}\!\!= \hbar (A_1 \! + \! A_2) g(\Omega) \coth\Big(\frac{\hbar \Omega}{2k_B T_N} \Big)
\int_{\q} \IM \chi^R_{\q \Omega}, \label{eq:SneqSP}
\end{eqnarray}
where $g(\Omega)$ denotes the spectrum of ferromagnetic resonance given by
\begin{eqnarray}
&& g(\Omega) = 2S_0^2 \Big( \frac{\gamma h_{\rm ac}}{2} \Big)^2 \frac{1}{(\Omega -\omega_0)^2 + \alpha^2 \Omega^2}.
\end{eqnarray}

\begin{figure}[!hbtp]
	\begin{center}
   \includegraphics[scale=0.55]{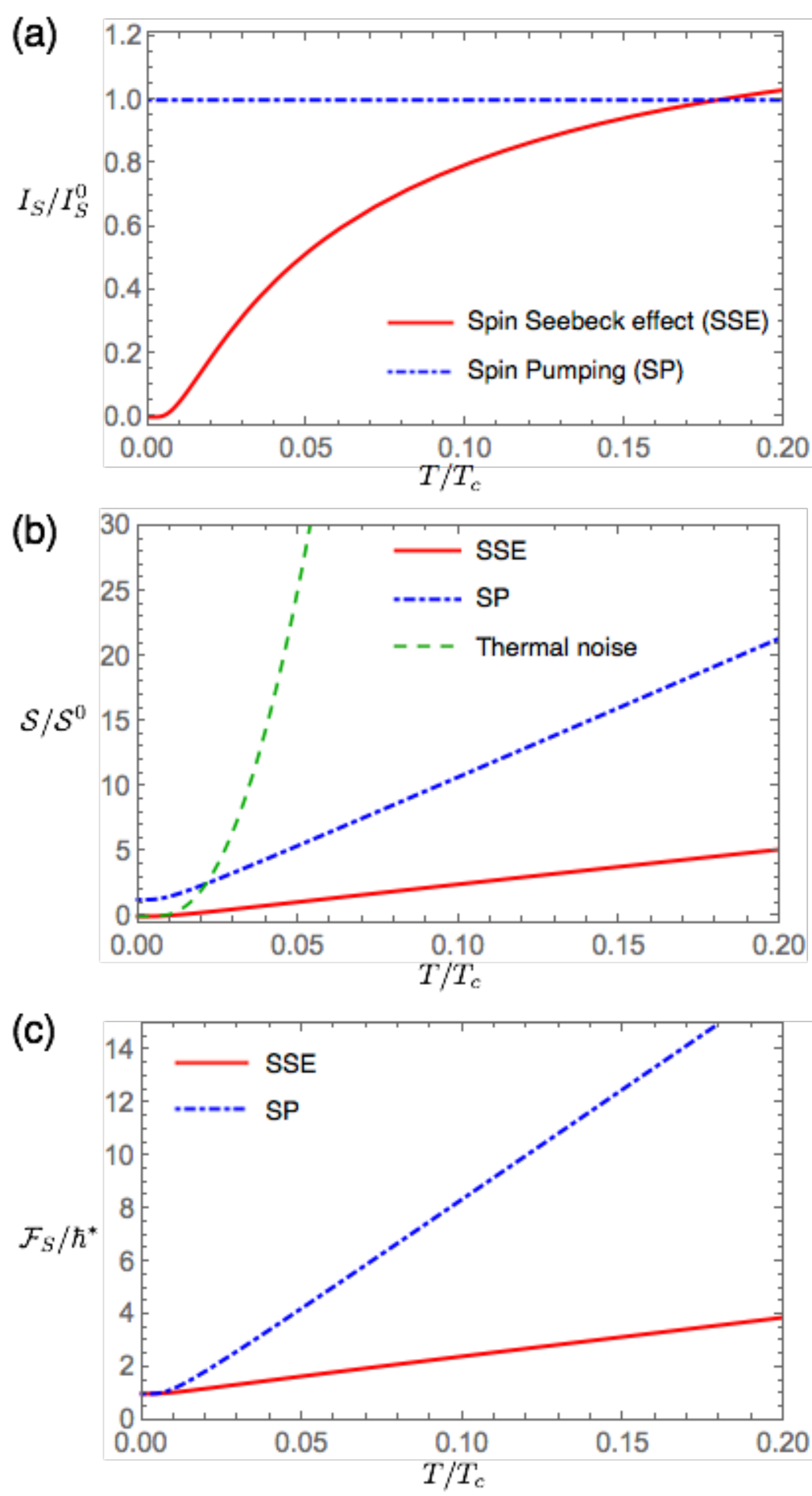}
	\caption{(Color online) Temperature dependence of (a) spin currents, 
    (b) nonequilibrium spin-current shot noises, and 
    (c) their ratios. The solid and dot-dashed lines indicate 
    the result of the spin Seebeck effect for the temperature bias $\Delta T=$1K 
    and the spin pumping, respectively. 
    In figure (b), thermal noise is denoted by the dashed line.
    }\label{fig_SPvsSSE}
	\end{center}
\end{figure}

\paragraph{Temperature dependence.---}
The spin Seebeck effect and spin pumping differ in terms of the nonequilibrium distribution difference $\delta f^{{\rm neq}}_{\r t\k\q\omega}$.
To elucidate the difference between the two mechanisms of spin current generation, we estimated spin currents and spin-current noises for a realistic situation by using the parameters of the YIG/Pt system in Ref.~\onlinecite{Kajiwara10} as the spin diffusion time of Pt $\tsf^{\rm Pt} = 0.3$ ps, Gilbert damping constant $\alpha = 6.7\times10^{-5}$, spin size $S_0=16$, and Curie temperature $T_c = 560$K. 
In addition, we assumed that the temperature bias at the interface is $\Delta T =1$K. 
Figure~\ref{fig_SPvsSSE}~(a) shows the estimated spin current as a function of temperature.
The plotted spin currents are normalized by the spin-pumping current at $T=0$K denoted as $I_S^0$. 
While the spin pumping current is almost independent of temperature, 
the spin Seebeck effect increases with temperature.
Figure~\ref{fig_SPvsSSE}~(b) shows the spin--current noises estimated using the same parameters. 
These noises were normalized against the spin pumping noise at $T=0$K, $\mathcal{S}^0$.  
In Fig.~\ref{fig_SPvsSSE}~(b), we show the thermal noise $\mathcal{S}^{\rm eq}$ 
by a dashed line. The temperature should be lowered sufficiently 
for accurate measurement of the nonequilibrium spin-current noises so that
the thermal noise is well suppressed.

\paragraph{Effective spin and statistics of magnons.---}
The ratio between the spin--current noise and the spin current, ${\cal S}/I_S$,
is calculated for spin Seebeck effect and spin pumping as
\begin{eqnarray}
& & \!\!\!\! {\cal F}_S^{\rm SSE} \! \equiv \! \frac{\SSSE }{ \IsSSE }  \! =\!
\frac{ \hbar^*\! \int_{\k\q\omega} \! \!\IM \chi^R_{\q \omega} \IM G^R_{\k\omega} (1\! +\! 2f^N_\omega )\frac{\del f^0_\omega}{\del T} }{ \int_{\k\q\omega} \IM \chi^R_{\q \omega} \IM G^R_{\k\omega} \frac{\del f^0_\omega}{\del T} }, \\
\label{S-Is-SSE}
& & \!\!\!\! {\cal F}_S^{\rm SP} \! \equiv \! \frac{\SSP }{ \IsSP } \! =\! \hbar^*
\coth \Big( \frac{\hbar \Omega}{2k_B T_N} \Big),
\end{eqnarray}
respectively, where
\begin{eqnarray}
\hbar^* = \hbar \frac{A_1 \! + \! A_2}{A_1 \! - \! A_2}.
\end{eqnarray}
Figure~\ref{fig_SPvsSSE}~(c) shows the temperature dependence of these ratios determined 
using the parameters estimated in the previous paragraph.
At low temperatures, the ratio approaches a constant value $\hbar^*$
for both spin Seebeck effect and spin pumping, which is interpreted 
as the effective spin carried by a magnon in analogy to
the effective charge of quasi-particles in current noise measurement~\cite{Kamra16a,Kamra16b}.
The effective magnon spin $\hbar^*$ is now determined by 
the ratio of the strengths of the spin--conserving process ($A_1$) 
and the spin--non-conserving process ($A_2$), and is enhanced from $\hbar$ in general.
This enhancement of the effective magnon spin originates from the mixture of two exchange processes
at the interface (see Supplementary Material for details).
At high temperatures, the ratio becomes proportional to the temperature 
for both mechanisms of spin-current generation. 
This result originates from the factor $(1+2f_\omega^N)$ in Eq.~(\ref{eq:Sneq}), which
represents a characteristic of the boson statistics of magnons~\cite{Aftergood17}.

\paragraph{Heating by microwave irradiation.---}
To describe spin pumping in a realistic situation, heating at the interface by microwave irradiation should be considered. 
Let us consider a spin pumping experiment, where the measured spin current $I_S^{\rm tot}$ consists of the spin current due to spin pumping $I_S^{\rm SP}$ and that due to heating $I_S^{\rm SSE}$ as $I_S^{\rm tot} = I_S^{\rm SP} + I_S^{\rm SSE}$. 
Similarly, the measured spin-current noise is given by $\mathcal{S}^{\rm tot} = \mathcal{S}^{\rm eq} + \mathcal{S}^{\rm SP} + \mathcal{S}^{\rm SSE}$.
Our aim here is to identify $I_S^{\rm SP}$ and $I_S^{\rm SSE}$ by measuring the spin current $I_S^{\rm tot}$, spin-current noise $\mathcal{S}^{\rm tot}$, and thermal noise $\mathcal{S}^{\rm eq}$, to determine the temperature bias due to the heating $\Delta T$, which cannot be measured directly. 
The spin currents $I_S^{\rm SP}$ and $I_S^{\rm SSE}$ can be rewritten by
\begin{eqnarray}
\vvv{\IsSP}{\IsSSE}\!=\!\mat{1/\FSP}{1/\FSSE}{1}{1}^{\!\! -1}\! \vvv{\Istot}{\Stot - \Seq }. \label{IsSSE-deltaS}
\end{eqnarray} 
By comparing Eq. (\ref{IsSSE-deltaS}) with Eq. (\ref{IsSSE}), we obtain the temperature bias due to the microwave irradiation at the interface $\Delta T$ as
\begin{eqnarray}
\Delta T =  \frac{  (\FSP - \FSSE )^{-1} (\Stot - \Seq + \FSP\Istot )}{   (A_1\! -\! A_2) \!\! \ikqom \IM \chi^{R}_{\q\omega} \IM G^{R}_{\k\omega} \frac{\del f^0_\omega}{\del T}  }.
\end{eqnarray}
Thus, the heating effect at the interface can be discussed by using the nonequilibrium 
spin-current noises.

\paragraph{Spin Hall angle.---}
If the spin--non-conserving process can be neglected ($A_2=0$), the ratio 
${\cal S}/I_S$ becomes a universal value $\hbar$, reflecting
the magnetization carried by one magnon.
For such a case, we can utilize the universal value of ${\cal S}/I_S$
as a standard for determining the conversion coefficient between
the spin current and the inverse spin Hall current of the NM, that is, the spin Hall angle.
The inverse spin Hall current induced by the spin current $I_S$ 
at the interface is expressed as $I_C^{\rm ISHE}= \theta_{\rm SH}^{-1} I_S$ 
with the spin Hall angle $\theta_{\rm SH}$.
Then, the inverse spin Hall current noise, $\mathcal{S}_C^{\rm ISHE}$, is written as
$\mathcal{S}_C^{\rm ISHE}=\theta_{\rm SH}^{-2}\mathcal{S}$.
By combining these relationships, the spin Hall angle is written as follows:
\begin{eqnarray}
\theta_{\rm SH} = \frac{\mathcal{S}/I_S}{\mathcal{S}_C^{\rm ISHE}/I_C^{\rm ISHE}}.
\label{SpinHallAngle}
\end{eqnarray}
If the value of $\mathcal{S}/I_S$ is known in advance, the spin Hall angle $\theta_{\rm SH}$
is determined by measuring $\mathcal{S}_C^{\rm ISHE}/I_C^{\rm ISHE}$.

\paragraph{Conclusion.---} 
In this study, we have investigated a spin-current noise at a FI--NM interface based on Keldysh Green's function. 
Using a general microscopic model, we have derived expressions for the spin current and the spin-current noise through the interface. 
The temperature dependence of both the spin Seebeck effect and spin pumping has been estimated using realistic experimental parameters for a YIG/Pt system. 
The spin--current noise contains useful information about spin transport.
We have demonstrated that simultaneous measurement of the spin current and the spin-current noise provides important information on effective magnon spin, heating effect under spin pumping, and the spin Hall angle of NMs.
Detailed analysis of the temperature dependence of the spin-current noise will be presented elsewhere.
We hope that the present calculation serves as a bridge between two well-established research areas, mesoscopic physics and spintronic physics.

\paragraph{Acknowledgements.---}
The authors are grateful to S. Takei, Y. Niimi, K. Kobayashi, and T. Arakawa for useful discussions and comments. 
This work is financially supported by ERATO-JST (JPMJER1402), and KAKENHI (Nos. 26103006, JP15K05153, JP15K05124, JP26220711, JP16H04023, JP26247063, and JP17H02927) from MEXT, Japan.



\begin{thebibliography}{10}
\bibitem{footnote1} 
In previous literatures (e.g. Ref.~\onlinecite{Du17}), the nonequilibrium distribution function is described by effective parameters such as a magnon chemical potential.
In the present calculation, we explicitly treat the nonequilibrium distribution funcion without using effective parameters.
\end{thebibliography}
\end{document}